# ФИЗИЧЕСКИЕ АСПЕКТЫ ПРОЦЕССОВ САМООРГАНИЗАЦИИ В КОМПОЗИТАХ

## 2. СТРУКТУРА И ВЗАИМОДЕЙСТВИЕ ИНТЕРЬЕРНЫХ ГРАНИЦ

Герега А.Н.

*Одесская государственная академия строительства и архитектуры, Украина*

**РЕЗЮМЕ**

Исследована модель осцилляторной составляющей взаимодействия интерьерных границ, рассмотрены особенности формирования структуры композитов в промежуточной асимптотике. В модели полимасштабной сети интерьерных границ получены аналитические выражения для расчета силовых полей предфрактала Серпинского и его модификаций.

**Ключевые слова:** структура композитов; интерьерные границы; силовое поле; ковер Серпинского; осцилляторное взаимодействие; фрактальные структуры; динамическая система; странные аттракторы; детерминированный хаос

# PHYSICAL ASPECTS OF THE PROCESSES SELF-ORGANIZATION IN COMPOSITES

## 2. STRUCTURE AND INTERACTION OF INNER BOUNDARIES

Herega A.N.

*Odessa State Academy of Civil Engineering and Architecture, Ukraine*

**SUMMARY**

Investigated the peculiarities of formation of composite structure in mesoscopic asymptotics, and the model of oscillatory component interaction inner boundaries. At the fractal model of the inner boundaries networks analytically determined the force field created by an arbitrary pre-fractal Sierpinski and its modification.

## 1. ВВЕДЕНИЕ

Интерьерные границы, как известно, является атрибутивной составляющей композиционных материалов. Их включение в состав структурных параметров материала имеет несколько причин [1, 2]. Во-первых, материал – сложная система, и, следовательно, должен иметь внутренние границы, существование которых является общесистемной закономерностью [3,4]. Во-вторых, образование внутренних границ следует рассматривать как механизм реализации одной из целевых функций системы – сохранения целостности: появление поверхностей раздела в материале, как известно, нивелирует избыточные напряжения. Кроме того, интерьерные границы являются неизбежным следствием процессов самоорганизации физического тела, обусловлены произвольностью формы кластеров и макроскопических структурных блоков [5].

## 2. ПОЛИМАСШТАБНОСТЬ СТРУКТУРЫ
## СЕТЕЙ ИНТЕРЬЕРНЫХ ГРАНИЦ

Поля механических напряжений, возникающих в твердом теле, как известно, существенно зависят от конфигурации неоднородностей. В случае квазилинейных внутренних границ, (которые интересуют нас в первую очередь), значения компонентов тензоров деформаций пропорциональны $r^{-1}$ [6], и, следовательно, на сравнительно больших расстояниях, многократно превышающих межатомные, их действие может быть существенным. Когда степень неоднородности материала достаточно возрастает, и во взаимодействии структурных элементов появляются коллективные эффекты, – возникает качественно новый этап его эволюции.

Ориентация линейных неоднородностей, возникающих на ранней стадии образования материала, конечно, неслучайны, но и не скоррелированы. По мере роста их плотности в ориентации вновь формирующихся линейных дефектов возникают преимущественные направления: поля деформаций линейных неоднородностей локально активизируют генерацию параллельных им дефектов, и в наименьшей степени препятствуют росту дефектов в перпендикулярном направлении. Таким образом, возникает характерная пространственная закономерность – самоаффинный мультифрактальный «узор» внутренних границ (рис.1), простейшими аналогами которого на плоскости и в объёме соответственно, могут быть модифицированные фракталы типа ковра Серпинского, губки Менгера и их дополнений (рис. 2 и 3) [7, 8]. Существенно, что суперпозицией необходимого количества модификаций этих предфракталов произвольных поколений можно восстановить «рисунок» сетей внутренних границ с наперед заданной точностью по аналогии с тем, как любая имеющая практическое значение функция может быть разложена в ряд Фурье. Кроме того, они с очевидностью позволяют учесть и такой существенный аспект процесса как множественность взаимодействующих очагов образования новых структурных блоков.

Статистическое самоподобие в расположении границ приводит (в результате интерференции) к некой «кратности» в конфигурации полей, к возникновению в локальных областях физического тела энергетических предпосылок для образования более крупных границ, к образованию конфигураций, играющих роль исходных элементов для структуры более высокого уровня [9,10]. В свою очередь, поля деформаций бόльших границ, воздействуя на границы меньших размеров, провоцируют их дальнейшее развитие. Это происходит синхронно во всех масштабах, и является, по сути, осцилляторным взаимодействием между разномасштабными внутренними границами [2].

## 3. СТОХАСТИЧЕСКАЯ МОДЕЛЬ ОСЦИЛЛЯТОРНОЙ СОСТАВЛЯЮЩЕЙ
## ВЗАИМОДЕЙСТВИЯ РАЗНОМАСШТАБНЫХ КВАЗИЛИНЕЙНЫХ
## СТРУКТУР

Материал и внутренние границы – взаимообусловленные и совместно развивающиеся кластерные системы [2,11]. Перераспределяя деформации в материале, ВГ эволюционируют, изменяя характерные размеры и осваивая новые

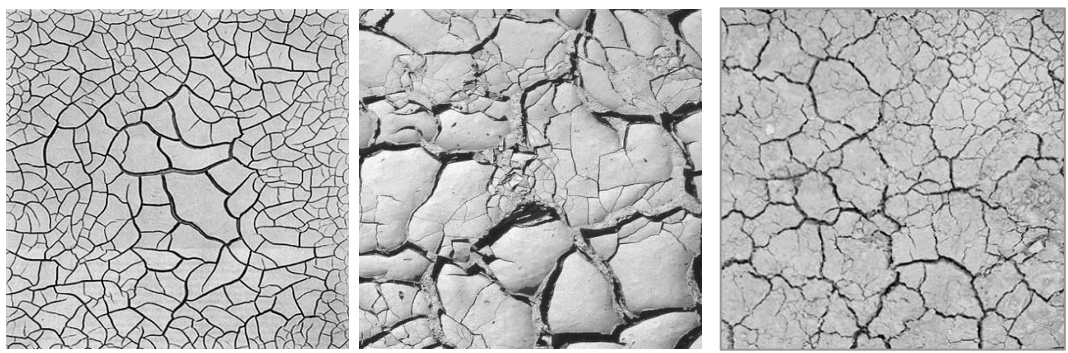

Рис. 1. Поверхностные трещины.

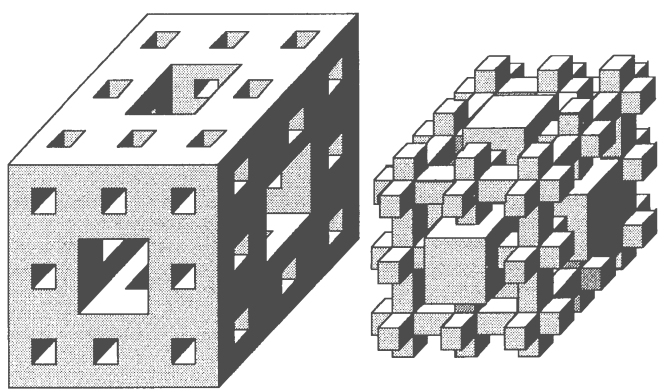

Рис. 2. Губка Менгера и её дополнение [10].

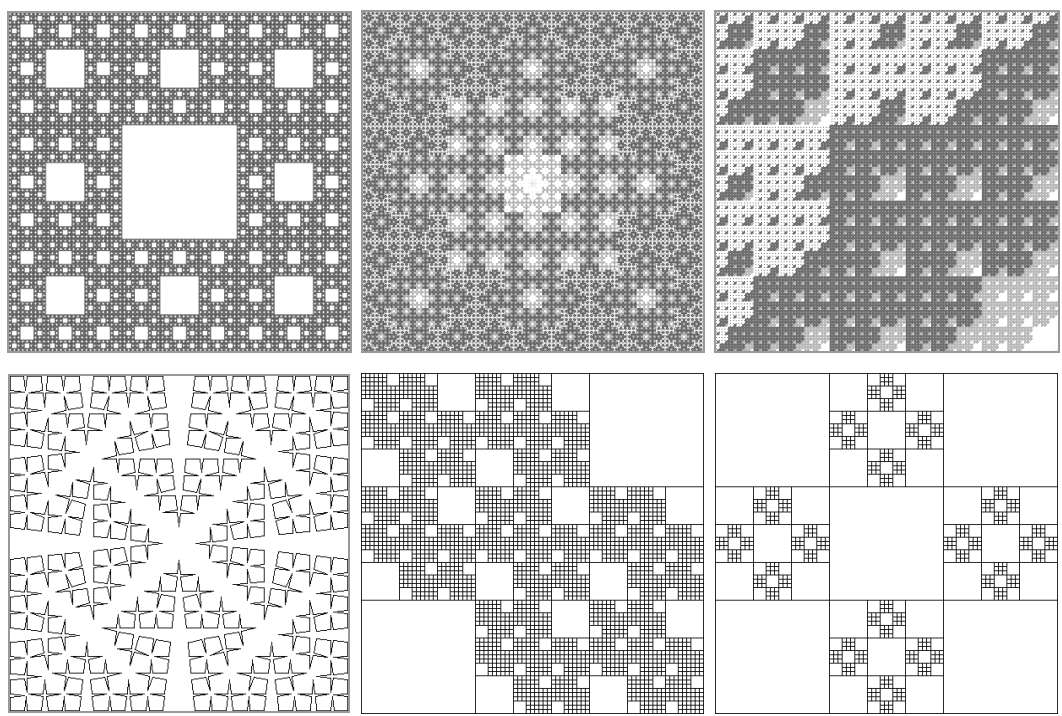

Рис. 3. «Вариации на тему» ковра Серпинского.

масштабы, тем самым, модифицируя материал.

Пусть физическое тело представляет собой автономную распределённую неконсервативную колебательную систему. Если предположить, что в системе действует обобщённая сила сопротивления, пропорциональная скорость распространения энергии ~~в системе~~ между линейными неоднородностями разных масштабных уровней, то в линейном приближении уравнение движения будет иметь стандартный вид

$$x'' + 2\gamma x' + \omega^2 x = 0,$$

где $\gamma$ – обобщённый коэффициент затухания, $\omega$ – циклическая частота.

Оценим условный период T таких колебаний [12] для системы линейных дефектов в прямоугольном параллелепипеде.

Введём величину, обратную коэффициенту жёсткости тела – податливость $C = 1/k$, которая определяет сколь далеко по масштабам может распространиться процесс трещинообразования, и для которой существует расчётная формула [13]

$$C = 8a^3/(Eh^3 b),$$

где $k$ – коэффициент жёсткости, $a$ – длина внутренней границы, $h$ – расстояние он границы до края тела, $b$ – толщина пластины, $E$ – модуль Юнга.

Предполагая $\gamma$ малым, имеем

$$\omega^2 \approx \omega_0^2 = 1/(m \cdot C),$$

тогда

$$T = 4\pi (2a^3 m / E h^3 b)^{1/2},$$

где $m$ – масса тела.

Другую оценку условного периода можно получить, определив логарифмический декремент затухания через последовательные (с интервалом в период) значения энергии системы $W_n$ [12, 14],

$$T = (1/\gamma) \cdot (W_n - W_{n+1}) / (W_n + W_{n+1}).$$

В работах [15-17] представлена математическая модель эволюции абстрактных систем, которые состоят из взаимодействующих по произвольным законам частей. Модель сформулирована как универсальная, и ее возможности продемонстрированы на описании процессов самоорганизации, возникающих в сепарирующихся циклических двухфазных потоках.

Рассмотрим процесс эволюции разномасштабных внутренних границ в гетерогенном материале как результат их взаимодействия.

Пусть структура материала представляет собой открытую динамическую систему. Рассмотрим три взаимодействующих масштабных уровня неоднородностей, и положим, что их эволюция описывается системой билинейных итерационных уравнений [2, 15-17]

$$\begin{cases} x_{n+1} = x_n - (k_{xy} + k_{low}) p x_n^2 + k_{yx} q y_n^2 + x_{in} \\ y_{n+1} = y_n + k_{xy} p x_n^2 - (k_{yx} + k_{yz}) q y_n^2 + k_{zy} r z_n^2 \\ z_{n+1} = z_n + k_{yz} q y_n^2 - (k_{zy} + k_{over}) r z_n^2 \end{cases} \quad (1)$$

где $x, y, z$ – динамические переменные, определяющие потенциальную энергию интерьерных границ определенного масштабного уровня, а $x_{in}$ – энергию внешнего воздействия; коэффициенты $k_{low}$ и $k_{over}$ задают долю рассеиваемой, $k_{ij}$ – долю переходящей между неоднородностями разных масштабов энергии, а коэффициенты $p, q, r$ – долю утилизируемой для перестройки энергии, причём, $\{k_{ij}\}$ и $\{p, q, r\} \in (0,1)$, $\{x, y, z\} \in R$.

Система уравнений в общем виде не интегрируется, но имеет стационарное решение, получаемое аналитически:

$$\begin{cases} x_{st} = \sqrt{\dfrac{x_{in}}{\left[k_{xy} + k_{low}\left(\dfrac{k_{yx}}{k_{yz}}\left(\dfrac{k_{zy}}{k_{over}}+1\right)+1\right)\right]p^{\left(\dfrac{k_{yx}}{k_{yz}}\left(\dfrac{k_{zy}}{k_{over}}+1\right)+1\right)}}} \\ y_{st} = \sqrt{\dfrac{x_{in} - k_{low} p \cdot x_{st}^2}{k_{yz} q}\left(\dfrac{k_{zy}}{k_{over}}+1\right)} \\ z_{st} = \sqrt{\dfrac{x_{in} - k_{low} p \cdot x_{st}^2}{k_{over} r}} \end{cases}$$

С ростом $x_{in}$ возможны два варианта эволюции системы. В первом – возникает каскад бифуркаций удвоений периода, и реализуется сценарий Фейгенбаума [18, 19] перехода к хаосу. При этом в фазовом пространстве потенциальных энергий масштабных уровней наблюдаются аттракторы, состоящие из $2^n$ точек. Во втором случае, после периодического режима возникает ситуация, аналогичная бифуркации Хопфа [19], приводящей, как известно, к возникновению квазипериодического режима. Дальнейшее увеличение $x_{in}$ приводит к бифуркации, в результате которой наступает хаотический режим, приводящий к возникновению в фазовом пространстве странного аттрактора [15,16]. На рис. 4 – примеры аттракторов, возникающих в системе.

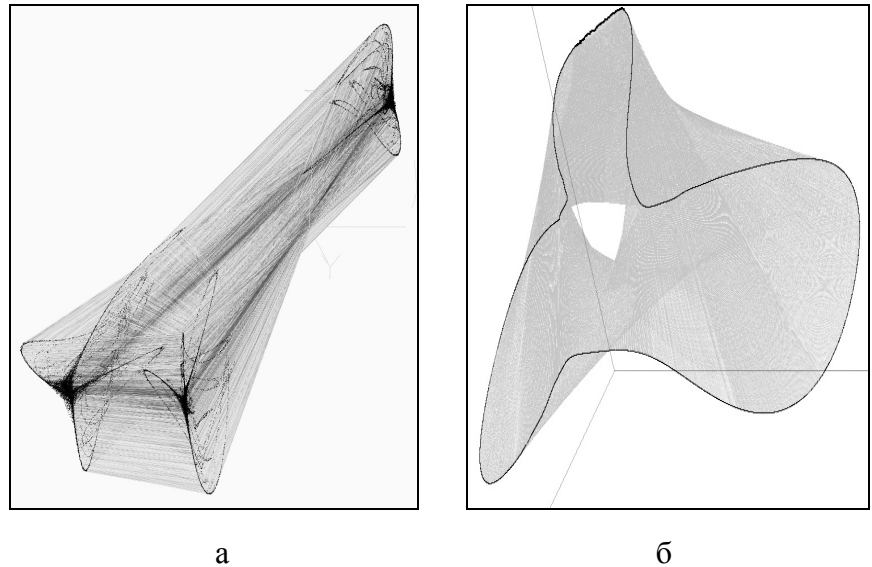

а                                б

Рис. 4. Странные аттракторы режима с перемежаемостью:
а – $k_{xy}=0.5$, $k_{yx}=0.4$, $k_{yz}=0.3$, $k_{zy}=0.3$, $k_{low}=0.7$, $k_{over}=0.4$, $p=1$, $q=1$, $r=1$, $x_{in}=1.8$;
б – $k_{xy}=0.5$, $k_{yx}=0.4$, $k_{yz}=0.3$, $k_{zy}=0.3$, $k_{low}=0.65$, $k_{over}=0.4$, $p=1$, $q=1$, $r=1$, $x_{in}=1.8365$.

Помимо рассмотренных сценариев развития хаоса в системе могут наблюдаться их комбинации, когда хаотическое поведение системы на какой-то промежуток времени сменяется периодическим.

Как известно, странные аттракторы является стохастическими автоколебаниями, поддерживаемыми в динамической системе за счёт внешнего источника.

Возникающие в фазовом пространстве энергий, в котором исследуется система (1), аттракторы можно интерпретировать как фигуры Лиссажу квазиколебательных процессов, что открывает возможность визуального определения характера и особенностей взаимодействия разномасштабных неоднородностей.

Коэффициенты в системе уравнений характеризуют особенности строения материала. В основу определения их численных значений в первом приближении могут быть положены максимально общие предположения, основанные на анализе физической ситуации. Определение этих значений для конкретных композиционных материалов видится на пути использования знаний о структуре и свойствах силовых полей внутренних границ [17].

## 4. МОДЕЛЬ СИЛОВОГО ПОЛЯ ПОЛИМАСШТАБНОЙ СЕТИ ИНТЕРЬЕРНЫХ ГРАНИЦ

Положим, что прямолинейные отрезки квадрата Серпинского соответствуют квазилинейным внутренним границам материала. Определим аналитически силовые поля, создаваемые полимасштабными сетями внутренних границ предфракталов Серпинского трёх типов, отличающихся симметрией, на произвольном шаге разбиения $m$.

Рассмотрим «проволочную» модель ковра Серпинского [17]. Пусть исходная квадратная рамка разделена четырьмя «проволоками» на девять равных квадратов. Процедура многократно повторяется на каждой из $8^m$ получаемых на очередном шаге рамок (за исключением центральных). Пусть также на каждой образующей рамок любого «поколения» с линейной плотностью $\lambda$ содержатся точечные источники, создающие поля с напряженностью $E \sim 1/r^2$.

Пусть ковёр Серпинского с длиной стороны образующего квадрата, равной $2H$, расположен так, что его центр совпадает с началом координат, а стороны параллельны осям. Составляющие вектора напряжённости, создаваемой отрезком, определятся соотношениями

$$\begin{cases} E_x = \lambda\,(\sin\alpha_2 - \sin\alpha_1)/r, \\ E_y = \lambda\,(\cos\alpha_2 - \cos\alpha_1)/r, \end{cases}$$

где $\alpha_i$ – угол между перпендикуляром длиной r, опущенным из точки, в которой определяется напряжённость, на отрезок или его продолжение, и соответствующим направлением на концевые точки отрезка.

Обозначим
$$A(u,v) = (u^2 + v^2)^{-1/2},\ B(u,v) = v/[u\,(u^2 + v^2)^{1/2}],$$

$$\xi(n,p) = -\xi + (-1)^n p,\ \eta(n,p) = -\eta + (-1)^n p,$$

тогда составляющие вектора напряжённости, создаваемой ковром $m$-го поколения в произвольных точках, не лежащих на прямых, которые содержат отрезки сети, можно вычислить по рекуррентным соотношениям

$$E_x = X_m(\xi, \eta) = \sum_{i=1}^{2} \{X_{m-1}(\xi, \eta(i, H'')) + X_{m-1}(\xi(i, H''), \eta) +$$

$$+ \sum_{j=1}^{2} [X_{m-1}(\xi(j, H''), y(i, H'')) + \lambda(-1)^j\,(A(\xi(j, H'), \eta(i, H')) -$$

$$- A(\xi(j, H), \eta(i, H')) + B(\xi(i, H'), \eta(j, H) - B(\xi(i, H'), \eta(j, H'))]\}$$

$$E_y = Y_m(\xi, \eta) = X_m(\eta, \xi),$$

где $H' = H/3$, $H'' = 2H/3$,

$$X_0(\xi, \eta) = \lambda \sum_{j=1}^{2} \sum_{i=1}^{2} [(-1)^i A(\xi(i, H), \eta(j, H)) + (-1)^{j+1} B(\xi(i, H), \eta(j, H))]$$

$$Y_0(\xi, \eta) = X_0(\eta, \xi) = \lambda \sum_{j=1}^{2} \sum_{i=1}^{2} [(-1)^i A(\eta(i, H), \xi(j, H)) +$$
$$+ (-1)^{j+1} B(\eta(i, H), \xi(j, H))].$$

Полученные соотношения – модификация результатов, описанных в [17].

Рассмотрим аналог ковра Серпинского с одной осью симметрии второго порядка (рис. 5).

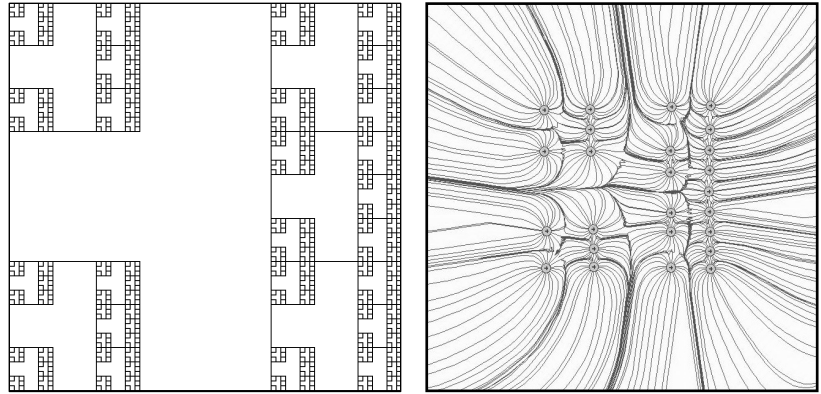

Рис. 5. Модификация ковра Серпинского и его силовое поле.

С использованием тех же обозначений, и с учетом $h = H/3^m$, поле такого ковра определяется соотношениями

$$X_n(\xi;\eta) = X_{n-1}(\xi - 2h_{n-1};\eta) + \sum_{i=1}^{2}\{(-1)^{i+1}B(\xi + h_n;\eta(i,h_{n-1})) +$$
$$+ \sum_{j=1}^{2}[X_{n-1}(\xi(i,2h_{n-1});\eta(j,2h_{n-1})) +$$
$$+(-1)^i\{A(\xi(i,h_{n-1});\eta(j,h_n)) - A(\xi(i,h_{n-1}) - 2h_{n-1};\eta(j,h_{n-1}))\}]\};$$

$$Y_n(\xi;\eta) = Y_{n-1}(\xi - 2h_{n-1};\eta) + \sum_{i=1}^{2}\{(-1)^i A(\xi + h_n;\eta(i,h_{n-1})) +$$
$$+ \sum_{j=1}^{2}[Y_{n-1}(\xi(i,2h_{n-1});\eta(j,2h_{n-1})) +$$
$$+(-1)^i\{B(\eta(j,h_{n-1});\xi(i,h_{n-1}) - 2h_{n-1}) - B(\eta(j,h_n);\xi(i,h_{n-1}))\}]\}.$$

Поле несимметричного ковра (рис. 6) может быть рассчитано по рекуррентным соотношениям:

$$X_1(\xi;\eta) = \sum_{i=1}^{2}\sum_{j=1}^{2}\sum_{l=1}^{2}(-1)^i[A(\xi(i,h_1);\eta(j,(2l-1)h_1/3)) -$$

$$-B(\xi(j,(2l-1)h_1/3;\eta(i,h_1))]; \quad Y_1(\xi;\eta) = X_1(\eta;\xi);$$

$$X_n(\xi;\eta) = X_{n-1}(\xi+2h_{n-1};\eta-2h_{n-1}) + \sum_{i=1}^{2}\{X_{n-1}(\xi;\eta(i,h_{n-1})-h_{n-1}) +$$

$$+(-1)^i[A(\xi-h_n;\eta(i,2h_{n-1})-h_{n-1}) - B(\xi(i,2h_{n-1})-h_{n-1};\eta-h_n)] +$$

$$+\sum_{j=1}^{2}[X_{n-1}(\xi(i,2h_{n-1});\eta(j,h_{n-1})+h_{n-1}) + (-1)^i\{A(\xi(i,h_{n-1});\eta(j,h_{n-1})+2h_{n-1}) -$$

$$-A(\xi(i,2h_{n-1})+h_{n-1};\eta(i,h_{n-1})+((-1)^j-1)h_{n-1}) +$$

$$+B(\xi(i,h_{n-1})+((-1)^j-1)h_{n-1};\eta(i,h_{n-1}))\}]\};$$

$$Y_n(\xi;\eta) = Y_{n-1}(\xi+2h_{n-1};\eta-2h_{n-1}) + \sum_{i=1}^{2}\{Y_{n-1}(\xi;\eta(i,h_{n-1})-h_{n-1}) +$$

$$+(-1)^i[A(\xi(i,2h_{n-1})-h_{n-1};\eta-h_n) - B(\eta(i,2h_{n-1})-h_{n-1};\xi-h_n)] +$$

$$+\sum_{j=1}^{2}[Y_{n-1}(\xi(i,2h_{n-1});\eta(j,h_{n-1})+h_{n-1}) + (-1)^{i+1}\{B(\eta(j,h_{n-1})+2h_{n-1};\xi(i,h_{n-1})) -$$

$$-B(\eta(i,h_{n-1})+((-1)^j-1)h_{n-1};\xi(i,2h_{n-1})+h_{n-1}) +$$

$$+A(\xi(i,h_{n-1})+((-1)^j-1)h_{n-1};\eta(i,h_{n-1}))\}]\}; \quad n=2,3...m;$$

$$E_x = E_x(x;y) = X_m(-x;-y), \quad E_y = E_y(x;y) = Y_m(-x;-y); \quad h_n = H \cdot 3^{n-m}.$$

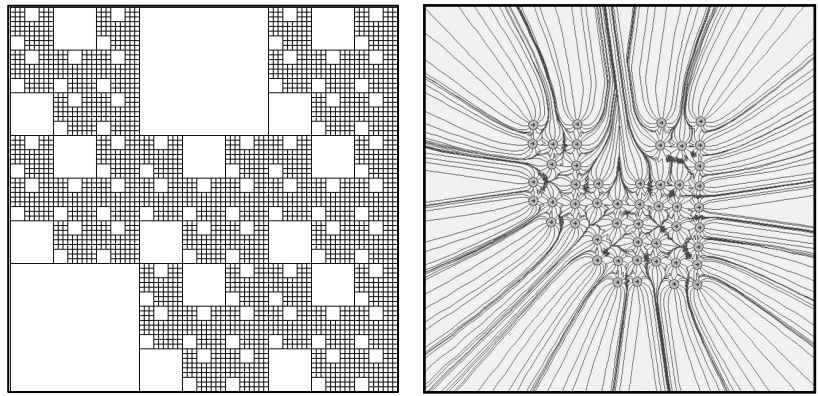

Рис.6. Несимметричная модификация ковра Серпинского.

## ЗАКЛЮЧЕНИЕ

Предложена математическая модель взаимодействия разномасштабных структур, позволяющая рассматривать композит в виде открытой нелинейной стохастической динамической системы. Модель сформулирована как универсальная, что даёт возможность исследовать развитие абстрактных систем, состоящих из подсистем, взаимодействующих по произвольным законам.

Проведен расчёт и визуализация силовых полей нескольких модификаций ковра Серпинского. Это позволяет, в соответствии с принципом суперпозиции, использовать регулярные фракталы для расчёта полей механических напряжений, создаваемых сетями внутренних границ произвольных конфигураций. Предметом отдельного исследования является возможность использования параметров полей для получения значений переходных коэффициентов, используемых в системе (1).

## ЛИТЕРАТУРА

1. Герега А.Н., Выровой В.Н. Интерьерные границы композитов: полимасштабность структуры и свойства силовых полей. / Труды IV симпозиума по механике композиционных материалов. – Москва: ИПРИМ РАН, 2012. – С. 204-209.
2. Выровой В.Н., Герега А.Н., Коробко О.А. Осцилляторное взаимодействие иерархически соподчинённых структур гетерогенного материала. / Труды конференции «Моделирование-2010». – Киев: ИПМЭ им. Г.Е. Пухова НАН Украины, 2010. – С. 253-260.
3. Уёмов А.И. Системы и системные параметры. / Проблемы формального анализа систем. – М.: Высшая школа, 1968. – 170 с.
4. Берталанфи Л. Общая теория систем: критический обзор. / Исследования по общей теории систем. – М.: Прогресс, 1969. – С. 23-82.
5. Соломатов В.И., Выровой В.Н., Бобрышев А.Н. Полиструктурная теория композиционных строительных материалов. – Ташкент: ФАН, 1991. – 345 с.
6. Косевич А.М. Физическая механика реальных кристаллов. – Киев: Наукова думка, 1981. – 328 с.
7. Божокин С.В., Паршин Д.А. Фракталы и мультифракталы. – Ижевск: НИЦ РХД, 2001. – 128 с.
8. Мандельброт Б. Фрактальная геометрия природы. – М.: ИКИ, 2002. – 656 с.
9. Панин В.Е., Гриняев Ю.В., Данилов В.И. Структурные уровни пластической деформации и разрушения. – Новосибирск: Наука, 1990. – 255 с.
10. Олемской А.И., Скляр И.А. Эволюция дефектной структуры твёрдого тела в процессе пластической деформации. // УФН. – 1992. – Т. 162, № 6. – С. 29-79.
11. Герега А.Н. Физические аспекты процессов самоорганизации в композитах. 1. Моделирование перколяционных кластеров фаз и внутренних границ. // Механика композиционных материалов и конструкций. – 2013. – Т.**. – №**. – С.**-**.
12. Мандельштам Л.И. Лекции по теории колебаний. – М.: Наука, 1972. – 470 с.
13. Броек Д. Основы механики разрушения. – М.: Высшая школа, 1980. – 368 с.
14. Андронов А.А., Витт А.А., Хайкин С.Э. Теория колебаний. – М.: Наука, 1981. – 568 с.
15. Герега А.Н., Лозовский Т.Л. Моделирование самоорганизации динамических дисперсных систем. I. Спонтанная организация двухфазного потока. // Электронное моделирование. – 2008. – Т. 30, № 3. – С. 3-12.
16. Bekker M., Herega A., Lozovskiy T. Strange Attractors and Chaotic Behavior of a Mathematical Model for a Centrifugal Filter with Feedback. // Advances in Dynamical Systems and Applications. – 2009. – Vol. 4, No. 2. – P. 179-194.
17. Герега А.Н., Дрик Н.Г., Угольников А.П. Ковер Серпинского с гибридной разветвленностью: перколяционный переход, критические показатели, силовое поле. // УФН. – 2012. – Т. 182, вып. 5. – С. 555-557.
18. Фейгенбаум М. Универсальность в поведении нелинейных систем. // УФН. – 1983. – Т. 141, вып. 2. – С. 343-374.
19. Берже П., Помо И., Видаль К. Порядок в хаосе. – М.: Мир, 1991. – 368 с.